\documentclass[%
 reprint,
 amsmath,amssymb,
 aps,
 prl,
superscriptaddress
]{revtex4-2}

\usepackage{graphicx}% Include figure files
\usepackage{dcolumn}% Align table columns on decimal point
\usepackage{bm}% bold math
\usepackage{xfrac}% for text fractions
\usepackage{enumitem}
\usepackage{natbib}

\begin{document}

\title{Quasiparticle interference observation of the topologically non-trivial drumhead surface state in ZrSiTe}

\author{B. A. Stuart}
\affiliation{Stewart Blusson Quantum Matter Institute, University of British Columbia, Vancouver, British Columbia, Canada V6T 1Z4}
\affiliation{Department of Physics and Astronomy, University of British Columbia, Vancouver, British Columbia, Canada V6T 1Z1}

\author{Seokhwan Choi}
\affiliation{Stewart Blusson Quantum Matter Institute, University of British Columbia, Vancouver, British Columbia, Canada V6T 1Z4}

\author{Jisun Kim}
\affiliation{Stewart Blusson Quantum Matter Institute, University of British Columbia, Vancouver, British Columbia, Canada V6T 1Z4}

\author{Lukas Muechler}
\affiliation{Center for Computational Quantum Physics, The Flatiron Institute, New York, New York, 10010, USA}

\author{Raquel Queiroz}
\affiliation{Department of Condensed Matter Physics, Weizmann Institute of Science, Rehovot 76100, Israel}

\author{Mohamed Oudah}
\affiliation{Stewart Blusson Quantum Matter Institute, University of British Columbia, Vancouver, British Columbia, Canada V6T 1Z4}
\affiliation{Department of Physics and Astronomy, University of British Columbia, Vancouver, British Columbia, Canada V6T 1Z1}
\affiliation{Department of Chemistry, Princeton University, Princeton, New Jersey 08544, USA}

\author{L. M. Schoop}
\affiliation{Department of Chemistry, Princeton University, Princeton, New Jersey 08544, USA}

\author{D. A. Bonn}
\affiliation{Stewart Blusson Quantum Matter Institute, University of British Columbia, Vancouver, British Columbia, Canada V6T 1Z4}
\affiliation{Department of Physics and Astronomy, University of British Columbia, Vancouver, British Columbia, Canada V6T 1Z1}

\author{S. A. Burke}
\affiliation{Stewart Blusson Quantum Matter Institute, University of British Columbia, Vancouver, British Columbia, Canada V6T 1Z4}
\affiliation{Department of Physics and Astronomy, University of British Columbia, Vancouver, British Columbia, Canada V6T 1Z1}
\affiliation{Department of Chemistry, University of British Columbia, Vancouver, British Columbia, Canada V6T 1Z1}

\date{\today}

\begin{abstract}
Drumhead surface states that link together loops of nodal lines arise in Dirac nodal-line semimetals as a consequence of the topologically non-trivial band crossings. We used low-temperature scanning tunneling microscopy and Fourier-transformed scanning tunneling spectroscopy to investigate the quasiparticle interference (QPI) properties of ZrSiTe. Our results show two scattering signals across portions of the drumhead state resolving the energy-momentum relationship through the occupied and unoccupied energy ranges it is predicted to span. Observation of this drumhead state is in contrast to previous studies on ZrSiS and ZrSiSe, where the QPI was dominated by topologically trivial bulk bands and surface states. Furthermore, we observe a near $\mathbf{k} \rightarrow -\mathbf{k}$ scattering process across the $\Gamma$-point, enabled by scattering between the spin-split drumhead bands in this material, showing the persistence of the drumhead state even in the presence of spin-orbit coupling.
\end{abstract}

\maketitle

\begin{figure*}
    \centering
    \includegraphics[width=\textwidth]{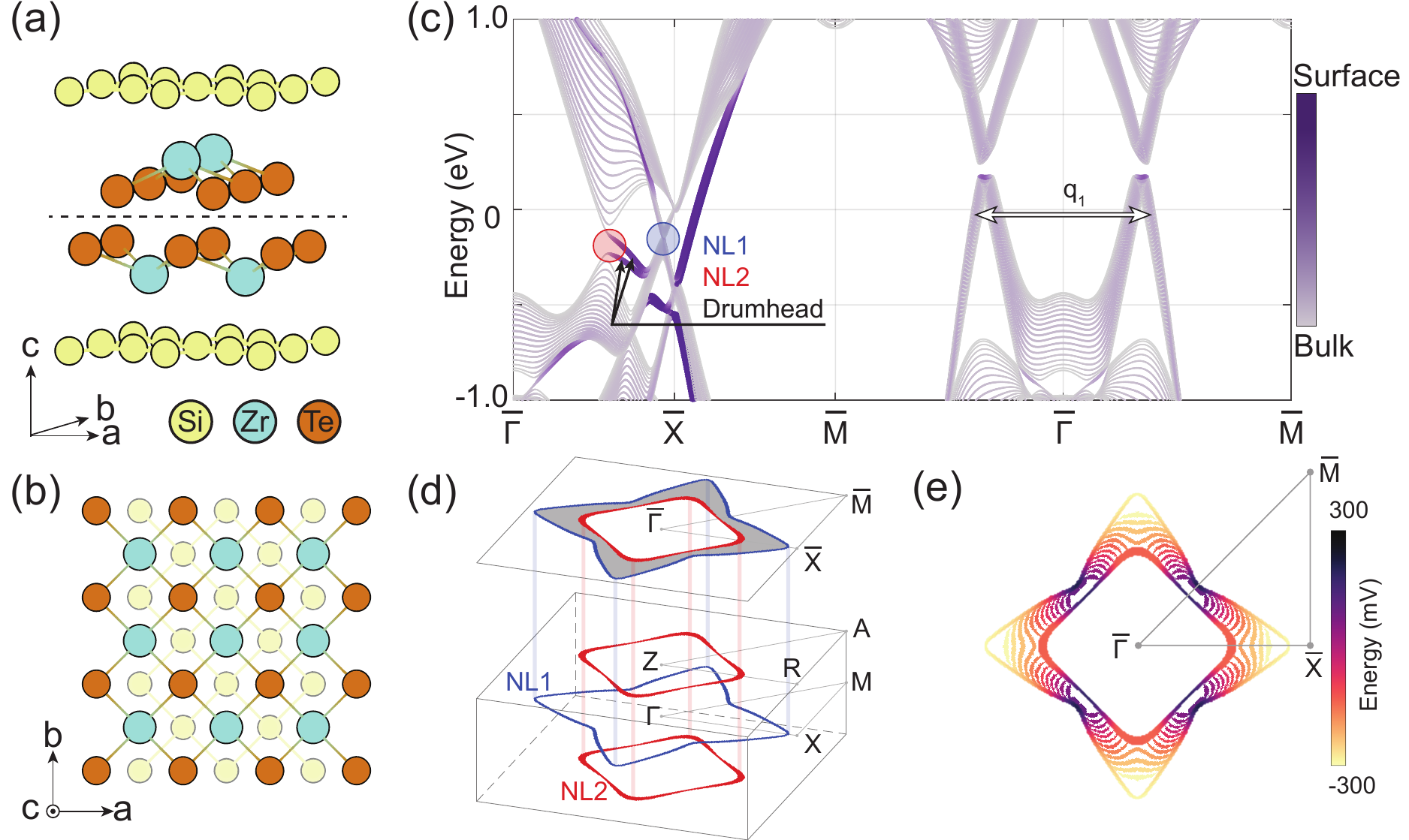}
    \caption{[Color online.] (a) ZrSiTe crystal structure. Dashed line indicates cleavage plane. (b) Cleaved ZrSiTe surface. Color faded Si atoms indicates they are below the ZrTe layer. (c) Band structure calculation with spin-orbit-coupling (SOC) (No SOC in SM Fig. 1) for an N = 20 unit cell thick slab along the [001] crystallographic direction. Two nodal lines are labeled in blue (NL1) and red (NL2), encircling the $\Gamma$-point and the Z-point, respectively. Black arrows indicate the drumhead states, splitting in energy due to spin-orbit coupling. (d) $k$-space structure of the Dirac nodal lines relevant in forming the drumhead state in the three-dimensional BZ and the [001] surface projected BZ. Shaded region of surface BZ indicates a non-trivial Berry phase resulting in drumhead states. (e) Energy landscape of NL1, NL2, and the drumhead states in the [001] surface projected BZ.}
    \label{fig:theory}
\end{figure*}

Topological semimetals (TSMs) are characterized by linearly dispersing band crossings in the bulk band structure that are protected by topological invariants and the symmetries of the material's crystalline space group. TSMs and their topologically protected surface states have garnered increasing attention in recent years, hosting a number of interesting electronic phenomena, including topologically robust boundary states and a linear non-saturating magnetoresistance \cite{Cd3As2_1}. The dimensionality and degeneracy of the linearly dispersing band crossings are used to classify the type of TSM, and are related to the resultant surface states \cite{SemimetalReview_1, SemimetalReview_2, SemimetalReview_3, SemimetalReview_4, SemimetalReview_5}. Materials which have band crossings in one-dimensional paths have been denoted topological nodal-line semimetals (NLSM), and when these paths form closed loops, topologically protected drumhead surface states arise. These span a two-dimensional region of the Brillouin zone (BZ) linking nodal loops together \cite{Drumhead_1, Drumhead_2, Drumhead_3, Drumhead_4, Drumhead_5, Drumhead_6, Drumhead_7, Drumhead_8, Drumhead_9, Drumhead_10}, similar to how Fermi-arcs in Weyl semimetals connect Weyl nodes \cite{TaAs_1, TaAs_2, TaAs_3, ChiralAnomaly_1}. To date, drumhead states have been sparsely studied experimentally mainly due to a lack of suitable candidate materials \cite{Drumhead_4, Drumhead_5, Drumhead_6, Drumhead_7, Drumhead_8, Drumhead_9, Drumhead_10}, and theoretical investigations often use drastically simplified models in superficial settings which do not necessarily transfer directly to real world materials. With drumhead states expected to exhibit some interesting electronic phenomena, for example a superconducting state with a transition temperature proportional to the drumhead size \cite{ref:drumhead_sc1, ref:drumhead_sc2}, finding new materials which host these topological surface states is of key research interest.

Non-symmorphic symmetries are known to allow, and in some cases enforce, bands to become degenerate along certain spaces within the BZ \cite{DSM_2D}, and non-symmorphic NLSMs were experimentally confirmed with the discovery of ZrSiS \cite{ZrSiS_ARPES1}. Since then, many NLSMs have been discovered among isostructural compounds sharing the form MXZ, (M = Zr, Hf), (X = Si, Ge, Sn), and (Z = O, S, Se, Te) \cite{SemimetalReview_4, SemimetalReview_5}. Among these compounds, most experimental focus to date has been placed on ZrSiS and ZrSiSe, particularly the topologically trivial features arising from the non-symmorphic degeneracies at the BZ boundaries and their resultant floating band surface states \cite{ZrSiS_MT1, ZrSiS_MT2, ZrSiS_MT3, ZrSiS_MT4, ZrSiS_MT5, ZrSiS_MT6, ZrSiS_MT7, ZrSiS_MT8, ZrSiS_MT9, ZrSiS_OC1, ZrSiS_UF1, ZrSiS_ARPES1, ZrSiS_ARPES2, ZrSiS_ARPES3, ZrSiS_ARPES4, ZrSiS_ARPES5, ZrSiS_ARPES6, ZrSiS_ARPES7, ZrSiS_STM_1, ZrSiS_STM_2, ZrSiS_RS1, ZrSiS_RS2, ZrSi(Se_Te)MT1, ZrSi(Se_Te)MT2, ZrSiSe_STM_1, ZrSiSe_STM_2}. This is due to a combination of the large contribution of these floating band surface states to the density of states (DOS) and the isolation of these features from other bands in the BZ. A difference in ZrSiTe is the emergence of a drumhead surface state spanning a large region of the BZ, which makes ZrSiTe the first material in the MXZ family suitable for studies of this state.

Recent angle-resolved photoemission spectroscopy (ARPES) measurements showed the first evidence of the drumhead surface state in ZrSiTe \cite{Drumhead_4}, resolving the occupied states. However, the drumhead state was predicted to extend over a wide energy range into the unoccupied region of the band structure which could not be measured. Scanning tunneling microscopy (STM) is a valuable and often complimentary tool to ARPES measurements due to its ability to probe states in both the occupied and unoccupied energy regions, along with providing spatial information of the material's surface. In this work, low-temperature STM and Fourier-transform scanning tunneling spectroscopy (FT-STS) were employed to investigate the scattering properties of the ZrSiTe drumhead state. The measurements track the dispersion of the drumhead state through the full energy domain and are in excellent agreement with theoretical calculations. We determine that the drumhead surface state provides the dominant contribution to the QPI, in contrast with ZrSiS and ZrSiSe, where the floating band surface state is most prominent. This marks the first time a real space technique has been used to observe the drumhead surface state, and shows that even in non-idealized real-world materials where many electronic bands compete and effects such as spin-orbit-coupling (SOC) are present, the drumhead state is robust and plays a significant role in the overall physics of ZrSiTe.

\begin{figure}
    \centering
    \includegraphics[width = \columnwidth]{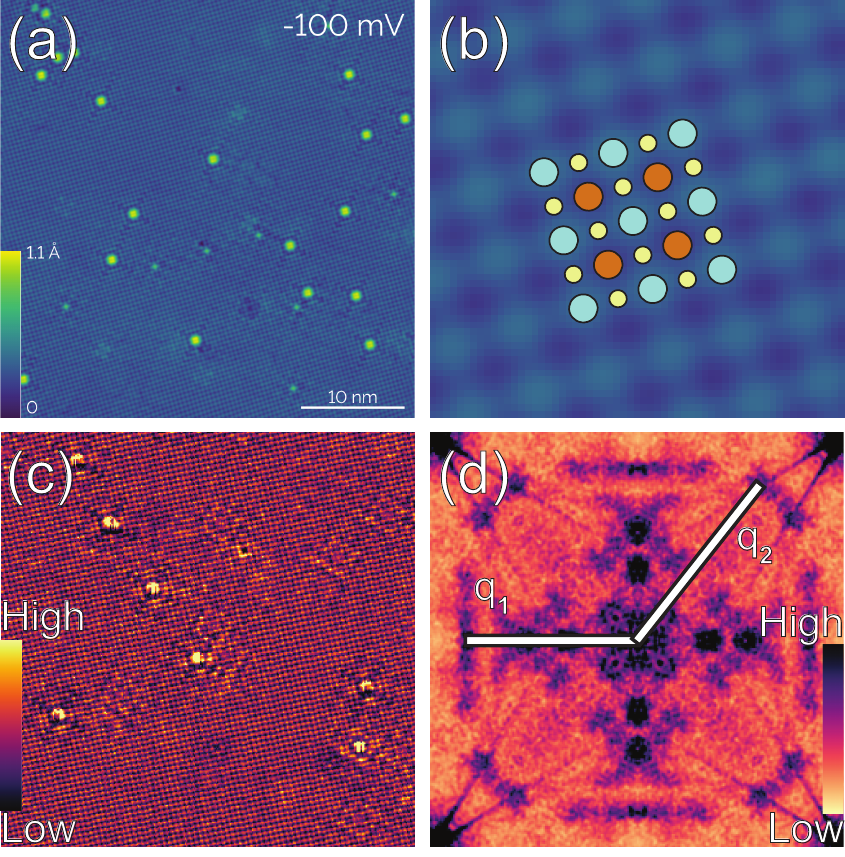}
    \caption{[Color online.] (a) $1024\times1024$ pixel, $40\times40$ nm$^2$ topographic image ($V_B$ = $-100$ mV, $I_t$ = $200$ pA). (b) Crystal structure overlay (same color scheme as Figure \ref{fig:theory}(a)). (c) Real space STS at a single energy. Oscillations around defects are clearly present, indicating QPI. (d) FT-STS at a single energy. Black corners of the image are the Bragg peaks. Scattering vectors shown are associated with the bulk Dirac nodal loop ($\mathbf{q_1}$), and the drumhead ($\mathbf{q_2}$).}
    \label{fig:stm}
\end{figure}

The ZrSiTe samples were grown using the techniques described in reference \cite{Topp_Lippmann_Varykhalov_Duppel_Lotsch_Ast_Schoop_2016}. The samples were cleaved \textit{in-situ} at room temperature at a pressure of 4 $\times$ 10$^{-10}$ mbar, then transferred into the STM (CreaTec GmbH). All measurements were performed at 4.5 K at a base pressure of $<10^{-10}$ mbar. Chemically etched tungsten tips were prepared \textit{in-situ} by e-beam heating and field emission. Imaging and STS were performed on gold prior to measurements on ZrSiTe to obtain a sharp, metallic tip.

Density functional theory (DFT) calculations were performed using the VASP package~\cite{VASP} with the standard pseudopotentials for Zr, Si, and Te. The experimental geometries were taken from the ICSD. For the self-consistent calculations, the reducible BZ was sampled by a $7\times7\times5$ k-mesh. A Wannier interpolation using 82 bands was performed by projecting onto an atomic-orbital basis centered at the atomic positions, consisting of Zr 5$s$,6$s$,5$p$,4$d$,5$d$, Si 3$s$,4$s$,3$p$,4$p$,3$d$ as well as Te 5$s$,6$s$,5$p$,6$p$,5$d$ orbitals. The theoretical spectra were calculated with an in-house code and the \textit{wanniertools}~\cite{wu2018wanniertools} package.

ZrSiTe crystallizes in the tetragonal P4/nmm space group (SG 129), consisting of Te-Zr-Si-Zr-Te quintuple layers held together weakly by van-der-Waals forces [Fig. \ref{fig:theory}(a)]. There is a natural cleavage plane between the quintuple layers which results in a surface consisting of a Te top layer, Zr hollow layer, and a Si square net layer \sfrac{1}{2} unit cell below the surface [Fig. \ref{fig:theory}(b)]. Of particular interest is the non-symmorphic mirror-glide-plane symmetry $ \bar{M_z} = \left\{ M_z \mid \sfrac{1}{2}, \sfrac{1}{2}, 0 \right\} $. This symmetry allows for bands to cross in the $k_z=0$ and $k_z=\sfrac{\pi}{c}$ planes protecting the nodal loops responsible for the formation of the drumhead surface state \cite{Topp_Lippmann_Varykhalov_Duppel_Lotsch_Ast_Schoop_2016, ZrSiS_ARPES6, Drumhead_4}.

DFT calculations were completed on a slab model consisting of N = 20 unit cells along the [001] crystallographic direction. Fig. \ref{fig:theory}(c) shows the resultant band structure. The nodal-line structure is presented in Fig. \ref{fig:theory}(d), isolating the Dirac crossings from other elements of the band structure. The two nodal loops relevant in the formation of the drumhead state are labeled NL1 and NL2 lying in the $k_z = 0$ and $k_z = \sfrac{\pi}{c}$ planes respectively, protected by the glide mirror symmetry $\bar{M_z}$.

\begin{figure*}
    \centering
    \includegraphics[width=\textwidth]{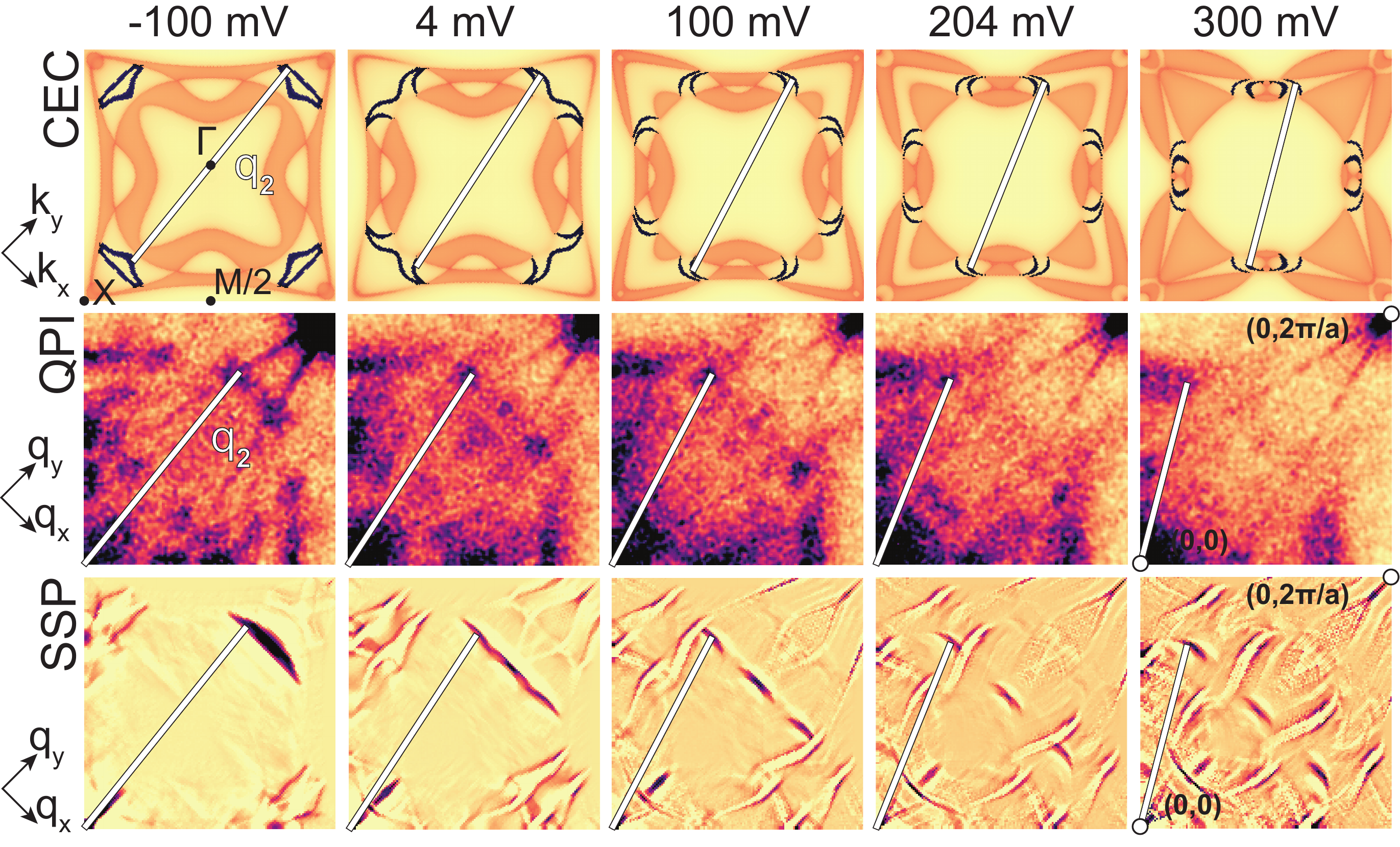}
    \caption{[Color online.] Top row: Constant energy contours (CECs) of the calculated surface band structure with SOC at the same energies as the QPI data (shifted by +165 mV as per [SM Fig. 2]). Highlighted in black are the drumhead states. $\mathbf{q_2}$ shows drumhead state scattering from next-nearest-neighbor drumhead states across the $\Gamma$-point.  Middle row: QPI data is 4-fold symmetrized, so only one quadrant of the QPI is shown, as other quadrants contain exactly the same information. \textbf{q} = (0,0) is positioned at the lower left corner of each QPI map, and the Bragg peak (either \textbf{q} = ($\pm 2\pi/a$,0) or \textbf{q} = (0,$\pm 2\pi/a$)) is positioned in the upper right corner. Bottom row: Spin-dependent scattering probability obtained through a spin-weighted autocorrelation of the CECs, showing an excellent agreement between QPI data and theory. The signal directly along the diagonal in all energies of the SSP calculations, and only at some energies in the QPI measurements, appears to be tip dependent, and is more prominent under certain measurement conditions. This is highlighted in the Supplementary Material [SM Fig. 4] which shows QPI data from three different measurements. Supplementary Material [SM Fig. 5] compares the evolution of the drumhead state scattering signal $q_2$ at all measured energies to the expected calculations on the same plot.}
    \label{fig:drumhead}
\end{figure*}

To explain the origin of the drumhead surface state, the Berry phase ($\gamma$) must be discussed in the frame of nodal loops projected onto the [001] surface BZ. This has been considered extensively in reference \cite{Drumhead_4}, where the momentum-resolved Berry phase of ZrSiTe has been investigated using Wilson loop analysis. In regions of the surface projected BZ where the Berry phase is an even multiple of $\pi$, the electronic states are topologically trivial. In contrast, any regions of the surface BZ where the Berry phase is an odd multiple of $\pi$, the electrons have a non-trivial topology. The area enclosed by a nodal loop projected onto the surface BZ carries with it a topologically non-trivial Berry phase of $\gamma = \pi$. This is additive with an increasing number of nodal loops, so in regions where two nodal loops overlap, the Berry phase is $\gamma = 2\pi$, which is again topologically trivial. From Fig. \ref{fig:theory}(d) it is clear that NL1 and NL2 have a significantly different $k$-space evolution from one another. When projected onto the surface BZ, this gives rise to a non-negligible two-dimensional area in the shape of a distorted annulus within which the Berry phase takes a value of $\gamma=\pi$ (gray area in Fig. \ref{fig:theory}(d)). This topological non-triviality leads to the formation of topologically protected drumhead surface states, linking the two nodal loops together in $k$-space \cite{Drumhead_4}.

Fig. \ref{fig:theory}(e) shows the calculated energy and momentum characteristics of the nodal loops and drumhead state on the surface projected BZ. From this perspective, it is clear that NL2 is fully enclosed within NL1. Within these nodal loops, several constant energy contours of the drumhead state are shown. These contours originate near the X point at approximately $-300$ mV, and evolve towards the $\Gamma$-M high symmetry line with increasing energy. At approximately the Fermi level, the drumhead contours break in two, forming crescent shapes as they continue to shift towards $\Gamma$-M. Upon reaching the $\Gamma$-M line, at approximately $300$ mV, the drumhead state terminates into the bulk nodal lines.

Fig. \ref{fig:stm}(a) shows a $40\times 40$ nm$^2$,  $1024 \times 1024$ pixel topographic image taken at a bias voltage of $V_B$ = $-100$ mV and a set point current of $I_t$ = $200$ pA. A clear atomic corrugation is present, with a lattice spacing of $4.02$ \AA. From a similar study on ZrSiS, the bright lattice seen can be attributed to the Zr atoms \cite{ZrSiS_STM_1}. Several different naturally occurring defects are present, and have a suitable concentration and spacing to observe strong QPI signals. Fig. \ref{fig:stm}(b) shows a magnified scan in a pristine area of the surface. The surface projected crystal structure overlaid shows the positions of Zr, Si, and Te atoms using the same color scheme as Fig. \ref{fig:theory}(a).

Scanning tunneling spectroscopy (STS) measurements were taken in the energy range of $-800$ mV to $800$ mV in 201 discrete steps. Fig. \ref{fig:stm}(c) shows a single energy slice of this STS measurement. The data was acquired over a 36 hour period using a $50\times 50$ nm$^2$, $512 \times 512$ pixel scan window with a set point of $500$ pA. Numerical differentiation was used to determine $dI/dV$. The raw current data was smoothed via a Gaussian filter using the same parameters that were required to achieve resolution of the surface state on the reference gold sample. The most intense QPI occurs around defects believed to be centered on Te sites, consistent with related studies \cite{ZrSiS_STM_1, ZrSiS_STM_2, ZrSiSe_STM_1, ZrSiSe_STM_2}.

Fig. \ref{fig:stm}(d) shows the QPI intensity map obtained by a Fourier transform of the STS. The QPI is symmetrized using the Bragg peaks as reference points, and further data processing is applied, including streak removal to account for minor tip changes during the measurement and defect masking to improve the resolution of small-$\mathbf{q}$ scattering vectors \cite{Shun_1}. All scattering signals present in the final QPI map are visible in the raw data, and can be seen alongside the data processing techniques in the Supplementary Materials [SM Fig. 3]. We concentrate on two scattering vectors in Fig. \ref{fig:stm}(d): $\mathbf{q_1}$ corresponds to quasiparticle scattering between the bulk hole-like (negative curvature) Dirac nodal loop bands across the $\Gamma$-point, along the $\Gamma$-M direction, indicated on the band structure calculation in Fig. \ref{fig:theory}(c), while $\mathbf{q_2}$ corresponds to scattering between the SOC split drumhead states. The drumhead state gives rise to the dominant surface scattering signals, contrary to ZrSiS and ZrSiSe where the floating band produces the dominant signal and any signature of the drumhead state has been absent. From $\mathbf{q_1}$ it was determined that a +165 mV energy shift applied to the theory was needed in order to align with the experimental data [SM Fig. 2], likely due to hole doping in the sample.

Fig. \ref{fig:drumhead} compares the calculated band structure to the drumhead state scattering: (top) the calculated constant energy contours (CECs), highlighting the drumhead state in black and shifting all bands by +165 mV as mentioned previously, (middle) the measured QPI data, symmetrized then quartered to remove redundant information and match the scale for the full BZ calculation, and (bottom) the spin-dependent scattering probability (SSP) determined by taking a spin-weighted autocorrelation of the CEC, shown at the same scale as the QPI maps. The data shows a distinct scattering signal linked to drumhead state scattering, $\mathbf{q_2}$, corresponding to scattering across the BZ between drumhead states near the X-points (through the $\Gamma$-point of the BZ)[Fig. \ref{fig:stm}(d) and Fig. \ref{fig:drumhead}]. While only $\mathbf{q_2}$ stands out in the data, the theoretical calculations show more possible scattering channels available than clearly visible in the QPI data, as is often the case, including the scattering vector connecting equivalent regions of the drumhead (diagonal) leading to the strong intensity between the two symmetric $\mathbf{q_2}$ regions in the SSP calculation. As different measurements show varying intensity in this region (see Supplementary Information), this discrepancy likely arises from tip or defect dependent sensitivity of the QPI to this particular scattering channel. 

%The $\mathbf{q_2}$ scattering vector becomes visible in the QPI at approximately -200 mV. As the energy is increased, the distance between these drumhead states in $k$-space is reduced, and consequently the length of $\mathbf{q_2}$ in the QPI is shortened in excellent agreement with the CECs. This behavior continues with increasing energy, and the scattering vector is shortened in a nearly linear trend up to approximately $300$ mV. At this point, the distance between the segments of the drumhead states that $\mathbf{q_2}$ connects becomes too small to resolve amongst the other small-$\mathbf{q}$ signals of the measurements mainly arising due to the defect structure of the crystal.

As energy is increased, the angle of $\mathbf{q_2}$ with respect to the $\Gamma$-M high-symmetry line varies. At around $-100$ mV, two QPI signals emerge near the Bragg peak. These two signals are indicative of the drumhead state breaking off into two separate branches. Nearing the Fermi level, this behavior becomes more obvious: the two peaks begin to move away from one another, flowing towards the $q_y$ = $q_x$ high-symmetry lines in the QPI maps. Into the unoccupied states above the Fermi level, the same trend is followed up to $300$ mV where the drumhead scattering signal overlaps with the signal from the bulk nodal line, and they become indistinguishable. The energy evolution of $\mathbf{q_2}$ is in excellent agreement with the calculated CECs. For a direct comparison over the full measured energy range, see Figure 5 in the Supplementary Information.

The presence of SOC leads to the splitting of the drumhead state, as discussed in \cite{Drumhead_4}, and we can ask if a split in the QPI is also possible. Although the resolution of this experiment is not sufficiently high to discern if the drumhead signal has a single or a double feature, we expect scattering between states of opposite (or close to opposite) spin to be drastically suppressed in the absence of both magnetic impurities and a magnetic tip (see e.g. the formalism in \cite{Queiroz2018}). In the particular case of the $\mathbf{q_2}$ scattering vector, which connects momenta close to $\mathbf{k}$ to $-\mathbf{k}$, the scattering between Kramer's pair states (exact back-scattering) is, in fact, fully suppressed due to the orthogonality of the Kramer's pair wavefunctions. Scattering between distinct drumhead states will be favored between those of almost aligned spin, and we expect that even in circumstances with increased resolution the single peak will dominate the QPI data.

In this study, we measured the surface of ZrSiTe using STM and through QPI we were able to visualize the evolution of the topologically protected drumhead surface state through its entire energy range, in both the occupied and unoccupied regions. Contrary to ZrSiS and ZrSiSe, where the floating band dominates the QPI, ZrSiTe exhibits a strong QPI signal from the drumhead state which is in excellent agreement with the theoretical calculations. Additionally, we showed that the nearby spin-split drumhead surface states allow for scattering across the BZ, in a near $\mathbf{k} \rightarrow -\mathbf{k}$ scattering process. As most current theoretical investigations into drumhead states consider highly idealized situations, we could not take for granted that experimental measurements would so closely mimic the calculations. The fact that we not only observe a scattering signal corresponding to the drumhead state, but that it dominates the physics even in a sea of competing electronic bands and presence of spin-orbit coupling, and is in exceptional agreement with the theory, is a great step towards understanding the fundamental electronic nature of these topologically protected states. We hope this clear evidence will open the door to further research into drumhead states, leading to the experimental confirmation of more host materials, as while the drumhead state has been observed with ARPES in other materials ( PbTeSe$_2$ \cite{Drumhead_5}, Co$_2$MnGa \cite{Drumhead_7}, and RAs$_3$ (R = Ca, Sr) \cite{Drumhead_10} to name a few), it has yet to be seen in any materials using STM outside of this work. Furthermore, the addition of STM and QPI to the suite of experimental probes used to study the topologically non-trivial drumhead states allows for a whole new and unexplored energy range that can be investigated, as the unoccupied states of materials are now accessible. We believe this could vastly expand upon the potential pool of materials to study, as now materials where drumhead states are predicted to appear above the Fermi level are no longer experimentally off limits.

Raw data and analysis code are available via the Open Science Framework (DOI: 10.17605/OSF.IO/7A8CU).

We would like to thank Yan Sun for providing the Wannier function fit. Work at UBC was supported by the Natural Sciences and Engineering Research Council Discovery Grant Program (RGPIN 2018-04271 and RGPIN-2018-04280) and Research Tools and Instruments (EQPEQ 473024-15), the Quantum Matter Institute, the Canada First Research Excellence Fund, the Canada Research Chairs program (S.B.), and the MP-UBC-UTokyo Center for Quantum Materials. Work at Princeton was supported by the Gordon and Betty Moore Foundation through Grant GBMF9064 to L.M.S. The Flatiron Institute is a division of the Simons Foundation. R. Q. was funded by the Deutsche Forschungsgemeinschaft (DFG, German Research Foundation) – Projektnummer 277101999 – TRR 183 (project B03), and the Israel Science Foundation.

\bibliography{DrumheadV2}

\end{document}